\documentclass[prl,aps,twocolumn,showpacs]{revtex4} 
\usepackage{graphicx} 
\usepackage{tabularx}
\usepackage{dcolumn} 
\usepackage{color}
\usepackage{bm}
\usepackage{amsmath}
\usepackage{amssymb}

\newcommand{\nn}{\nonumber}

\begin{document} 
 
\title{%
Scattering theory on surface Majorana fermions by an impurity in $^3$He-B}

\author{Yasumasa Tsutsumi} 
\affiliation{Department of Basic Science, University of Tokyo, Meguro, Tokyo 153-8902, Japan} 

\date{\today}

\begin{abstract}
We have formulated the scattering theory on Majorana fermions emerging in the surface bound state of the superfluid $^3$He B-phase ($^3$He-B) by an impurity.
By applying the theory to the electron bubble, which is regarded as the impurity, trapped below a free surface of $^3$He-B, the observed mobility of the electron bubble [Ikegami {\it et al.}, J. Phys. Soc. Jpn. {\bf 82}, 124607 (2013)] is quantitatively reproduced.
The mobility is suppressed in low temperatures from the expected value in the bulk $^3$He-B by the contribution from the surface Majorana fermions.
By contrast, the mobility does not depend on trapped depth of the electron bubble in spite of the spatial variation of the wave function of the surface Majorana fermions.
Our formulated theory demonstrates the depth independent mobility by considering intermediate states in the scattering process.
Therefore, we conclude that the experiment has succeeded in observing Majorana fermions in the surface bound state.
\end{abstract}

 
 
\maketitle 


{\it Introduction.---}
Detection of Majorana fermions in topological superfluids/superconductors~\cite{hasan:2010,qi:2011,mizushima:2015,mizushima:2016} is a modern objective in condensed matter physics.
Among several systems hosting Majorana fermions~\cite{fu:2008,schnyder:2008,sato:2009,lutchyn:2010,sasaki:2011,tsutsumi:2013,tsutsumi:2015,xu:2015,kawakami:2015}, the superfluid $^3$He B-phase ($^3$He-B) is an ideal one because the bulk properties have been well established~\cite{vollhardt:book,volovik:book}.
The established theory allows us quantitative comparison with experimental results.
In $^3$He-B, Majorana fermions emerge in the surface bound state with linear dispersion below the superfluid gap~\cite{chung:2009,nagato:2009,tsutsumi:2011b,tsutsumi:2012c,wu:2013}.

There have been a few reports on the successful detection of the Majorana fermions in $^3$He-B~\cite{okuda:2012,bunkov:arXiv}.
In an experiment, the impedance by the surface acoustic wave of $^3$He-B was measured in a sample cell~\cite{okuda:2012}.
The observed impedance {\it qualitatively} corresponds with the theoretically evaluated impedance by considering the surface bound state.
Another experiment demonstrates the enhancement of heat capacity of $^3$He-B in low temperatures~\cite{bunkov:arXiv}.
This enhancement is interpreted as caused by the contribution from the Majorana fermions in the surface bound state; however, the heat capacity is larger than an estimated value from the geometrical surface area of the sample owing to the surface roughness.
Thus, it is hard to {\it quantitatively} compare theoretical predictions and experimental results for $^3$He-B in a vessel owing to the surface roughness with the atomic scale.
At the rough surface, the Majorana fermions have a broad spectrum deviated from linear dispersion~\cite{murakawa:2009}.

The specular reflection of $^3$He quasiparticles (QPs) is realized on a free surface.
The experiment to observe the Majorana fermions in the surface bound state at the free surface of $^3$He-B was performed by Ikegami {\it et al.}~\cite{ikegami:2013b}.
In the experiment, mobility of electron bubbles trapped below the free surface is measured.
Note that the electron bubble is a self-trapped electron by the Pauli exclusion with electrons surrounding $^3$He atoms~\cite{ahonen:1978}.
Since the drag force on the slowly moving electron bubbles is caused by momentum transfer from the $^3$He QPs, the Majorana fermions in the surface bound state are expected to be observed via the mobility.
The formulation on the mobility of the electron bubbles in the bulk $^3$He is well established~\cite{baym:1977,baym:1979,salomaa:1980}, which reproduces experimental results~\cite{ahonen:1976,ahonen:1978,roach:1977}.
Furthermore, the transverse force on the moving electron bubbles in the superfluid $^3$He A-phase, which is the direct evidence of chiral symmetry breaking~\cite{ikegami:2013,ikegami:2015}, is quantitatively demonstrated by considering momentum transfer from Weyl fermions in the bound state around the electron bubble~\cite{shevtsov:2016}.

In the experiment by Ikegami {\it et al.}~\cite{ikegami:2013b}, the observed mobility is suppressed in low temperatures from the theoretically evaluated value for $^3$He-B in the bulk~\cite{baym:1977,baym:1979}.
The suppression of the mobility can be interpreted as caused by the scattering with low energy Majorana fermions.
However, the observed mobility is independent of the trapped depth of the electron bubbles in the whole measuring range $21\ {\rm nm}\le z\le 58\ {\rm nm}$ in spite of that local density of states (DOS) of the Majorana fermions is modulated with the scale of the coherence length, $\xi\sim 100$ nm.
Thus, it has been unclear on the origin of the suppressed mobility and whether the experiment succeeded in observing the Majorana fermions in the surface bound state.

In this Letter, we formulate the scattering between an impurity and the Majorana fermion in the surface bound state so as to understand the mobility observed by the above experiment~\cite{ikegami:2013b}.
The mobility obtained by our formulation quantitatively reproduces the observed depth independent mobility as a result of that the depth dependence of the scattering cross section and the local DOS of the surface bound state cancel out.

{\it Formulation.---}
The mobility of impurities, such as the electron bubbles, is determined by the momentum transfer from the $^3$He QPs to the impurity.
The equation of motion for the momentum of the impurity is given by~\cite{bromley:1981,shevtsov:2016}
\begin{align}
\frac{d{\bm P}}{dt}=-\sum_{{\bm k},{\bm k}',\sigma,\sigma'}\hbar({\bm k}'-{\bm k})(1-f_{{\bm k}'})f_{\bm k}\Gamma_{\bm v}({\bm k},\sigma\to{\bm k}',\sigma'),
\end{align}
where $f_{\bm k}=[1+\exp(E_{\bm k}/k_{\rm B}T)]^{-1}$ is the Fermi distribution at temperature $T$ for the QP with excitation energy $E_{\bm k}$ and $\Gamma_{\bm v}({\bm k},\sigma\to{\bm k}',\sigma')$ is the transition rate for the QPs from wave number ${\bm k}$ and spin $\sigma$ to ${\bm k}'$ and $\sigma'$ by the scattering on the moving impurity with velocity ${\bm v}$.
By the time reversal symmetry in $^3$He-B, the equation of motion becomes up to the first order of ${\bm v}$~\cite{bromley:1981,shevtsov:2016},
\begin{multline}
\frac{d{\bm P}}{dt}=-\frac{\hbar^2}{2k_{\rm B}T}\sum_{{\bm k},{\bm k}',\sigma,\sigma'}[{\bm v}\cdot({\bm k}'-{\bm k})]({\bm k}'-{\bm k})(1-f_{\bm k})f_{\bm k}\\
\times\Gamma({\bm k},\sigma\to{\bm k}',\sigma'),
\end{multline}
where $\Gamma$ is the transition rate for the QPs by the scattering on the static impurity:
\begin{align}
\Gamma({\bm k},\sigma\to{\bm k}',\sigma')=\frac{2\pi }{\hbar }\delta(E_{{\bm k}'}-E_{\bm k})|t({\bm k},\sigma\to{\bm k}',\sigma')|^2.
\end{align}
Here, we only consider the elastic scattering because the recoil energy of the impurity is sufficiently low according to the experiments~\cite{ahonen:1978,ahonen:1976,roach:1977}.
Since the drag force on the moving impurity with ${\bm v}_{\parallel }$ parallel to the free surface is given by $d{\bm P}/dt=-\eta_{\parallel }{\bm v}_{\parallel }$, the Stokes drag coefficient is obtained by
\begin{multline}
\eta_{\parallel }=\frac{\pi\hbar }{2}\sum_{{\bm k},{\bm k}'}({\bm k}_{\parallel }'-{\bm k}_{\parallel })^2\left(-\frac{\partial f_{\bm k}}{\partial E_{\bm k}}\right)\delta(E_{{\bm k}'}-E_{\bm k})\\
\times\sum_{\sigma,\sigma'}|t({\bm k},\sigma\to{\bm k}',\sigma')|^2.
\label{eq:eta1}
\end{multline}

The squared $T$-matrix element is given by
\begin{align}
\sum_{\sigma,\sigma'}|t({\bm k},\sigma\to{\bm k}',\sigma')|^2=\sum_{\sigma,\sigma'=\uparrow,\downarrow }|\langle\Psi_{{\bm k}',\sigma'}|T_{\rm S}|\Psi_{{\bm k},\sigma }\rangle|^2.
\end{align}
For the surface bound state in $^3$He-B, the spinors for the Majorana fermions are described by $|\Psi_{{\bm k},\uparrow }^{\pm }\rangle=\frac{1}{\sqrt{2}}e^{-i\phi/2}(1,0,0,i)^{\rm T}|{\bm k}\rangle$ and $|\Psi_{{\bm k},\downarrow }^{\pm }\rangle=\mp\frac{1}{\sqrt{2}}e^{i\phi/2}(0,i,-1,0)^{\rm T}|{\bm k}\rangle$ for the QP energy $E_{\bm k}=\pm\Delta\sin\theta$~\cite{chung:2009,nagato:2009,supplement1}, where $\Delta$ is the bulk gap and $\phi$ and $\theta$ are the azimuthal angle and the polar angle from the normal axis to the free surface, respectively.
The wave number of the Majorana fermions is fixed on the Fermi wave number, ${\bm k}=k_{\rm F}\hat{\bm k}$.
The $T$-matrix element $\langle{\bm k}'|T_{\rm S}|{\bm k}\rangle\equiv T_{\rm S}(\hat{\bm k}',\hat{\bm k},E,z)$ is given by the following equation based on the Lippman-Schwinger equation~\cite{salomaa:1980,shevtsov:2016}:
\begin{multline}
T_{\rm S}(\hat{\bm k}',\hat{\bm k},E,z)=T_{\rm N}(\hat{\bm k}',\hat{\bm k})+N_{\rm F}\int\frac{d\Omega_{{\bm k}''}}{4\pi }T_{\rm N}(\hat{\bm k}',\hat{\bm k}'')\\
\times\left[g_{\rm S}(\hat{\bm k}'',E,z)-g_{\rm N}\right]T_{\rm S}(\hat{\bm k}'',\hat{\bm k},E,z),
\label{eq:LS}
\end{multline}
where $N_{\rm F}$ is the DOS in the normal state per spin.
Since the size of the impurity is much less than the coherence length, the $T$-matrix depends on the quasiclassical Green's function $g_{\rm S}(\hat{\bm k},E,z)$ for the surface bound state only at the impurity position $z$~\cite{thuneberg:1981}.
The quasiclassical Green's function in the normal state is given by $g_{\rm N}=-i\pi\tau_0$ and the expression of $g_{\rm S}$ is obtained by Ref.~\onlinecite{tsutsumi:2012c} (see also Supplemental Material~\cite{supplement2}).
The $T$-matrix in the normal state is given by $T_{\rm N}(\hat{\bm k}',\hat{\bm k})={\rm diag}[t_{\rm N}(\hat{\bm k}',\hat{\bm k})\sigma_0,-t_{\rm N}(-\hat{\bm k}',-\hat{\bm k})^*\sigma_0]$ with
\begin{align}
t_{\rm N}(\hat{\bm k}',\hat{\bm k})=-\frac{1}{\pi N_{\rm F}}\sum_{l=0}^{\infty }(2l+1)e^{i\delta_l}\sin\delta_lP_l(\hat{\bm k}\cdot\hat{\bm k}'),
\end{align}
where $P_l$ is the Legendre polynomial and $\delta_l$ is the phase shift depending on the potential of the impurity.
Here, $\tau_0$ ($\sigma_0$) is the unit matrix in the Nambu- (spin-)space.
By solving Eq.~\eqref{eq:LS}, we derive the squared $T$-matrix element in Eq.~\eqref{eq:eta1}.

The summation of wave number for initial and final states in Eq.~\eqref{eq:eta1} can be replaced by the integral:
\begin{align}
\sum_{\bm k}\to\int_{-\Delta }^{\Delta }dE_{\bm k}\int\frac{d\Omega_{\bm k}}{4\pi }N(\hat{\bm k},E_{\bm k},z),
\label{eq:sumtoint}
\end{align}
where the angle-resolved local DOS of the surface bound state is given by~\cite{tsutsumi:2012c}
\begin{multline}
N(\hat{\bm k},E_{\bm k},z)=N_{\rm F}\frac{\pi\Delta|\cos\theta|}{4}{\rm sech}^2\left(\frac{z}{2\xi }\right)\\
\times[\delta(E_{\bm k}-\Delta\sin\theta)+\delta(E_{\bm k}+\Delta\sin\theta)],
\label{eq:DOS}
\end{multline}
where $\xi\equiv\hbar v_{\rm F}/2\Delta$ with the Fermi velocity $v_{\rm F}$.
Thus, the depth dependence of the impurity mobility originates from the DOS of the surface bound state and the $T$-matrix element through the quasiclassical Green's function.
By the replacement of the summation by the integral, the drag coefficient in Eq.~\eqref{eq:eta1} is reduced to
\begin{multline}
\eta_{\parallel }=\frac{\pi^2}{16}{\rm sech}^4\left(\frac{z}{2\xi }\right)n_3p_{\rm F}\int_{-\Delta }^{\Delta }dE\left(-\frac{\partial f}{\partial E}\right)\\
\times\frac{3}{2}\int_0^{2\pi }d\varphi(1-\cos\varphi)\overline{\frac{d\sigma }{d\Omega }}(\varphi,E,z),
\label{eq:eta2}
\end{multline}
where $n_3=k_{\rm F}^3/3\pi^2$ is $^3$He density, $p_{\rm F}$ is the Fermi momentum, and $\varphi\equiv\phi'-\phi$.
The polar angle averaged differential cross section $\overline{d\sigma/d\Omega }$ is given by
\begin{multline}
\overline{\frac{d\sigma }{d\Omega }}(\varphi,E,z)=\left(\frac{\pi N_{\rm F}}{k_{\rm F}}\right)^2\left(\frac{E}{\Delta }\right)^4\frac{1}{4}\sum_{s,s'=\pm 1}\\
\times\sum_{\sigma,\sigma'}|t({\bm k}_{\parallel },sk_{\perp },\sigma\to{\bm k}_{\parallel }',s'k_{\perp },\sigma')|^2,
\label{eq:dsdp}
\end{multline}
where the amplitude of the perpendicular wave number $k_{\perp }\equiv k_{\rm F}|\cos\theta|=\sqrt{k_{\rm F}^2-k_{\parallel }^2}$ is fixed by setting the QP energy and the sign of that only remains as degree of freedom in the $T$-matrix element.
The fourth power of the QP energy, $(E/\Delta)^4$, originates from the linear dispersion of the surface bound state, $(k_{\parallel }/k_{\rm F})^2=(E/\Delta)^2$, and the Jacobian $\sin\theta\sin\theta'$ being replaced by $(E/\Delta)^2$ owing to the delta function in the DOS of the surface bound state in Eq.~\eqref{eq:DOS}.
Here, we define the total cross section and transport cross section as
\begin{align}
\sigma_{\rm tot}(E,z)\equiv&\frac{3}{2}\int_0^{2\pi }d\varphi\overline{\frac{d\sigma }{d\Omega }}(\varphi,E,z),\\
\sigma_{\rm tr}(E,z)\equiv&\frac{3}{2}\int_0^{2\pi }d\varphi(1-\cos\varphi)\overline{\frac{d\sigma }{d\Omega }}(\varphi,E,z),
\end{align}
respectively.
Then, the drag force on the impurity moving parallel to the free surface from the surface Majorana fermions is obtained by the energy integral of the transport cross section.

So far, we generally formulate the scattering theory on the Majorana fermions in the surface bound state by an impurity with the following assumptions.
(I) The velocity of the moving impurity is sufficiently low, i.e., the drag force is independent of the impurity velocity.
(II) The recoil energy of the impurity is sufficiently low, i.e., the collision process is the elastic scattering.
(III) The size of the impurity is much less than the coherence length, i.e., the $T$-matrix depends on the quasiclassical Green's function only at the impurity position.
From now on, we consider the electron bubble as the impurity whose potential is well modeled by the hard sphere potential.
This model potential quantitatively reproduces not only the observed mobility in the A- and B-phases~\cite{baym:1977,baym:1979,salomaa:1980} but also the transverse force on the electron bubble in the A-phase caused by chiral symmetry breaking~\cite{shevtsov:2016,shevtsov:arXiv}.
For the hard sphere potential with radius $R$, the phase shift $\delta_l$ is given by $\tan\delta_l=j_l(k_{\rm F}R)/n_l(k_{\rm F}R)$, where $j_l$ and $n_l$ are the spherical Bessel and Neumann functions, respectively.
The only parameter, the radius $R$ of the hard sphere potential, is fixed by the observed mobility in the normal state $\mu_{\rm N}=1.8\times 10^{-6}\ {\rm m^2V^{-1}s^{-1}}$~\cite{ikegami:2013b} at $R=11.17k_{\rm F}^{-1}$~\cite{shevtsov:2016,shevtsov:arXiv}.

\begin{figure}
\begin{center}
\includegraphics[width=8.5cm]{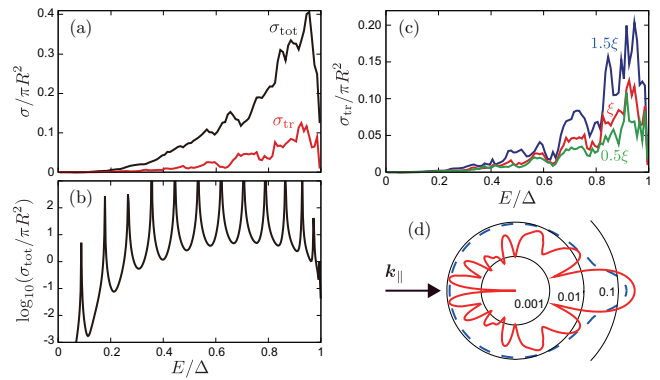}
\end{center}
\caption{\label{fig1}(Color online)
(a) Total cross section and transport cross section at $z=\xi$.
(b) Total cross section at $z=10\xi$.
(c) Transport cross sections at $z=0.5\xi$, $\xi$, and $1.5\xi$.
(d) Polar angle averaged differential cross section scaled by $\pi R^2$ at $z=\xi$ for $E=0.9\Delta$ (solid line) and that in the normal state scaled by $10\pi R^2$ (dashed line).
}
\end{figure}

{\it Scattering cross section.---}
First, we discuss the scattering cross section between the electron bubble and the Majorana fermion in the surface bound state.
In Fig.~\ref{fig1}(a), we show the total cross section $\sigma_{\rm tot}(E,z)$ and the transport cross section $\sigma_{\rm tr}(E,z)$ at $z=\xi$.
Here, the scattering cross sections have the relation, $\sigma(-E)=\sigma(E)$, by the symmetry of the Majorana fermions.
The cross sections have the peak structures on account of the quasi-bound state around the electron bubble which decays to the surface Majorana fermion with a finite life time.
The peak structures are clearly seen in the total cross section at $z=10\xi$ [Fig.~\ref{fig1}(b)] because the quasi-bound state has a long life time deep in $^3$He-B from the free surface.
Each peak is derived from each partial wave $l \lesssim k_{\rm F}R$.
When the electron bubble approaches to the surface, the peaks broaden out and the partial waves interfere with each other.
Note that the bound state around the electron bubble is never detected in the bulk because low energy QPs below the superfluid gap are absent.
The influence of the quasi-bound state emerges only after the presence of the low energy Majorana fermions in the surface bound state.

The transport cross section is strongly suppressed from the total cross section.
In order to clarify the reason of the suppression, the polar angle averaged differential cross section scaled by $\pi R^2$ at $z=\xi$ for $E=0.9\Delta$ is depicted in Fig.~\ref{fig1}(d) by a solid line.
The polar angle averaged differential cross section should be compared with that in the normal state depicted by a dashed line with the scale $10\pi R^2$, where the total cross section and transport cross section in the normal state are $\sigma_{\rm tot}^{\rm N}\approx 2\pi R^2$ and $\sigma_{\rm tr}^{\rm N}\approx \pi R^2$, respectively.
By the comparison between them, it is clear that the strong suppression of the transport cross section is due to the reduction of the backscattering.
The reduction of the backscattering with oscillation of the differential cross section results from constructive interference of the partial waves with $l\lesssim k_{\rm F}R$ in the forward direction and destructive interference of them at other angles as well as the bulk B-phase~\cite{baym:1979}.
In particular, the perfect backscattering of the Majorana fermion from momentum ${\bm k}_{\parallel }$ to $-{\bm k}_{\parallel }$ is forbidden.
It is similar to the edge state in topological insulators which is not scattered to backward by nonmagnetic impurities owing to the time-reversal symmetry~\cite{qi:2011}.
Note that the electron bubble is regarded as a nonmagnetic impurity for the $^3$He QPs with small nuclear magnetic moment.

The depth dependence of the transport cross section is shown in Fig.~\ref{fig1}(c).
As the electron bubble approaches to the free surface, the transport cross section becomes smaller.
This depth dependence is due to the life time of the quasi-bound state around the electron bubble, which is short near the free surface.
The life time is embedded in the quasiclassical Green's function $g_{\rm S}(\hat{\bm k},E,z)$~\cite{tsutsumi:2012c,supplement2} describing the intermediate states in the scattering process.

{\it Mobility.---}
The QP energy integral of the transport cross section in Eq.~\eqref{eq:eta2} provides the drag coefficient $\eta_{\parallel }$ on the electron bubble by the Majorana fermions in the surface bound state.
On mobility, however, the drag force also acts on the electron bubble by the QPs in the continuum state with the energy above the superfluid gap.
When we assume that the transport cross section for the QPs in the continuum state corresponds to that in the bulk $^3$He-B, $\sigma_{\rm tr}^{\rm B}(E)$, the drag coefficient is given by~\cite{baym:1977,baym:1979}
\begin{align}
\eta_{\rm B}=2n_3p_{\rm F}\int_{\Delta }^{\infty }dE\left(-\frac{\partial f}{\partial E}\right)\sigma_{\rm tr}^{\rm B}(E).
\end{align}
From the drag coefficients, we can obtain the mobility as $\mu=e/(\eta_{\parallel }+\eta_{\rm B})$, where $e$ is the electron charge.
The temperature dependence of the mobility comes from the amplitude of the gap $\Delta(T)$ and the derivative of the Fermi distribution $-\partial f/\partial E\propto{\rm sech}^2(E/2k_{\rm B}T)$.
In low temperatures, therefore, the mobility is dominated by the Majorana fermions in the surface bound state because the low energy QPs have a major contribution to the drag force.
Then, the assumption that the transport cross section in the continuum state is replaced by that in the bulk is relevant in low temperatures.

\begin{figure}
\begin{center}
\includegraphics[width=8.5cm]{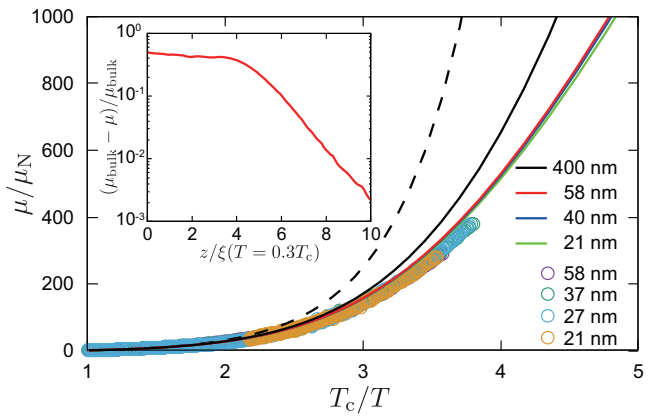}
\end{center}
\caption{\label{fig2}(Color online)
Temperature dependence of calculated mobility (solid lines) compared with the experimental result (circles) in Ref.~\onlinecite{ikegami:2013b}.
The solid lines indicate the mobility at $z=21, 40, 58$, and 400 nm from the small value.
The dashed line indicates the mobility in the bulk $^3$He-B.
The experimental data have error bars about $\pm 2\%$ of $\mu/\mu_{\rm N}$ which are smaller than the size of the data symbols.
Systematic error in temperature due to calibration is less than 3\%~\cite{ikegami:private}.
Inset: depth dependence of the calculated mobility at $T=0.3T_{\rm c}$.
}
\end{figure}

In Fig.~\ref{fig2}, we compare the temperature dependence of the calculated mobility (solid lines) and the experimental result (circles) in Ref.~\onlinecite{ikegami:2013b}.
The experimental data have error bars about $\pm 2\%$ of $\mu/\mu_{\rm N}$ which are smaller than the size of the data symbol.
Systematic error in temperature due to calibration is less than 3\%~\cite{ikegami:private}.
The calculated mobility is shown for positions of the electron bubble at $z=21, 40, 58$, and 400 nm.
The obtained mobility is strongly suppressed in low temperatures from the bulk value of the mobility (dashed line) and less sensitive to the position of the electron bubble between $z=21$ nm and 58 nm.
At $T=0.3T_{\rm c}$, $\mu(z=58\ {\rm nm})/\mu(z=21\ {\rm nm})\approx 1.03$.
The tiny depth dependence can be hardly observed by the experiment.
Although it seems that the value of the calculated mobility is slightly larger than the experimental value, their difference is smaller than the possible calibration error on the experimental temperature.
Thus, the calculated mobility quantitatively agrees with the experimental result.

Note that although the present experimental data were measured under a magnetic field $B=30$ mT perpendicular to the free surface, the experiment without magnetic field observed mobility with the same temperature dependence of them~\cite{ikegami:2013b}.
The magnetic field leads to a gap $\mu_{\rm n}B$ in the excitation energy of the surface bound state~\cite{nagato:2009}, where $\mu_{\rm n}$ is the nuclear magnetic moment of the $^3$He atom.
Since $\mu_{\rm n}B/\Delta\sim 0.01$ for $B=30$ mT, influence of the small gap on the mobility is not observed in the experimental temperature range owing to small DOS of the surface bound state in low energy.
The confirmation of the field independence of the mobility is an evidence of that the surface Majorana fermion has linear dispersion.
Field dependence of the mobility will be observed in low temperature below $k_{\rm B}T\sim\mu_{\rm n}B$.

The depth insensitive mobility results from the cancellation of the spatial variation of the squared local DOS of the surface bound state proportional to ${\rm sech}^4(z/2\xi)$ by the depth dependence of the transport cross section $\sigma_{\rm tr}(E,z)$ in Fig.~\ref{fig1}(c).
For the electron bubble far away from the free surface, since the transport cross section has the depth independent spectrum in Fig.~\ref{fig1}(b), the drag force $\eta_{\parallel }$ from the Majorana fermions decreases proportionally to the squared DOS and the mobility approaches to the bulk value, $\mu_{\rm bulk}$.
The depth dependence of $\mu_{\rm bulk}-\mu$ at $T=0.3T_{\rm c}$ is shown in the inset of Fig.~\ref{fig2}.
The difference exponentially decreases with the scale of the coherence length in the depth larger than $z=5\xi(T=0.3T_{\rm c})$.
Here, $\xi(T=0.3T_{\rm c})=\hbar v_{\rm F}/2\Delta(T=0.3T_{\rm c})\approx 138$ nm.

{\it Summary.---}
We have formulated the scattering theory on Majorana fermions in the surface bound state by an impurity in $^3$He-B.
By applying the theory to the electron bubble trapped below a free surface of $^3$He-B, we have quantitatively reproduced the observed depth independent mobility in Ref.~\onlinecite{ikegami:2013b}.
The depth independence is due to the cancellation of the DOS of the surface bound state and the suppression of the scattering cross section by the short life time of the quasi-bound state around the electron bubble.
Owing to the quantitative agreement on the electron bubble mobility by the experiment and this theory, we conclude that the experiment has succeeded in observing Majorana fermions in the surface bound state.
The detected Majorana fermions without magnetic field should have ideal linear dispersion due to the perfect agreement with the theoretical expectation.

I thank H.~Ikegami, K.~Kono, J.~A.~Sauls, Y.~Kato, and Y.~Masaki for helpful discussions.
I also acknowledge financial support from the Japan Society for the Promotion of Science (JSPS).
This work was supported by JSPS KAKENHI Grants No.~15K17715 and No.~15J05698.


%

\clearpage

\renewcommand{\thefigure}{S\arabic{figure}} 

\renewcommand{\thetable}{S\arabic{table}} 

\renewcommand{\thesection}{S\arabic{section}.}

\renewcommand{\theequation}{S.\arabic{equation}}

\setcounter{figure}{0}
\setcounter{table}{0}
\setcounter{equation}{0}

\begin{flushleft} 
{\Large {\bf Supplemental Material}}
\end{flushleft} 

\section{Majorana fermions in surface bound state}

In the superfluid state with an inhomogeneous order parameter, the Bogoliubov wave function $\Psi({\bm r})=(u_{\uparrow }({\bm r}),u_{\downarrow }({\bm r}),v_{\downarrow }({\bm r}),v_{\uparrow }({\bm r}))^{\rm T}$ obeys the Bogoliubov--de Gennes equation~\cite{skawakami:2011}:
\begin{align}
\int d{\bm r}_2
\begin{pmatrix}
\epsilon({\bm r}_1,{\bm r}_2) & -\Delta({\bm r}_1,{\bm r}_2) \\
-\Delta^{\dagger }({\bm r}_2,{\bm r}_1) & -\epsilon^{\rm T}({\bm r}_2,{\bm r}_1)
\end{pmatrix}
\Psi({\bm r}_2)=E\Psi({\bm r}_1).
\end{align}
For the superfluid $^3$He B-phase ($^3$He-B), the normal state Hamiltonian is $\epsilon({\bm r}_1,{\bm r}_2)=\delta({\bm r}_1-{\bm r}_2)[\hbar^2(-\nabla^2-k_{\rm F}^2)/2m]\sigma_0$, where $k_{\rm F}$ is the Fermi wave number, $m$ is the effective mass of $^3$He, and $\sigma_0$ is the unit matrix in the spin-space.
The pair potential is
\begin{align}
\Delta({\bm r}_1,{\bm r}_2)=\int\frac{d{\bm k}}{(2\pi)^3}\Delta({\bm r},{\bm k})e^{i{\bm k}\cdot{\bm r}'},
\end{align}
with ${\bm r}=({\bm r}_1+{\bm r}_2)/2$ and ${\bm r}'={\bm r}_1-{\bm r}_2$.
When $^3$He-B fills $z>0$ and there is a specular surface at $z=0$, the spatial dependence of the pair potential is described by $\Delta({\bm r},{\bm k})=i\frac{1}{k_{\rm F}}[\Delta_{\parallel }(z)(k_x\sigma_x+k_y\sigma_y)+\Delta_{\perp }(z)k_z\sigma_z]\sigma_y$ with the parallel component $\Delta_{\parallel }(z)$ and the perpendicular component $\Delta_{\perp }(z)$ to the surface, where ${\bm \sigma }$ is the Pauli matrix in the spin-space.
By using the expression of $\Delta({\bm r},{\bm k})$, the pair potential is given by~\cite{smatsumoto:2001}
\begin{align}
\Delta({\bm r}_1,{\bm r}_2)=&-\int\frac{d{\bm k}}{(2\pi)^3}\frac{1}{k_{\rm F}}[\Delta_{\parallel }(z)(\partial_{x_2}\sigma_x+\partial_{y_2}\sigma_y)\nn\\
&+\Delta_{\perp }(z)\partial_{z_2}\sigma_z]\sigma_ye^{i{\bm k}\cdot({\bm r}_1-{\bm r}_2)}\nn\\
=&-\frac{1}{k_{\rm F}}[\Delta_{\parallel }(z)(\partial_{x_2}\sigma_x+\partial_{y_2}\sigma_y)+\Delta_{\perp }(z)\partial_{z_2}\sigma_z]\sigma_y\nn\\
&\times\delta({\bm r}_1-{\bm r}_2).
\end{align}
Therefore, the Bogoliubov-de Gennes equation is reduced to
\begin{align}
\begin{pmatrix}
\epsilon({\bm r}) & -\mathcal{D}({\bm r}) \\
\mathcal{D}^{\dagger }({\bm r}) & -\epsilon^{\rm T}({\bm r})
\end{pmatrix}
\Psi({\bm r})=E\Psi({\bm r}),
\end{align}
where $\epsilon({\bm r})=\hbar^2(-\nabla^2-k_{\rm F}^2)/2m\sigma_0\equiv\epsilon_0({\bm r})\sigma_0$, and
\begin{align}
\mathcal{D}({\bm r})=&\frac{1}{k_{\rm F}}[\Delta_{\parallel }(z)(\partial_x\sigma_x+\partial_y\sigma_y)+\Delta_{\perp }(z)\partial_z\sigma_z]\sigma_y\nn\\
&+\frac{1}{2k_{\rm F}}\frac{\partial\Delta_{\perp }(z)}{\partial z}\sigma_z\sigma_y.
\end{align}
Within the Andreev approximation, $\epsilon_0({\bm r})\approx\hbar(-i{\bm \nabla }-{\bm k}_{\rm F})\cdot{\bm v}_{\rm F}$ with the Fermi velocity ${\bm v}_{\rm F}$ and the last term of $\mathcal{D}({\bm r})$ is negligible because $\Delta_{\perp }(z)$ varies with the scale of the coherence length.

Here, we assume that only the perpendicular component $\Delta_{\perp }(z)$ is suppressed at the specular edge as $\Delta_{\perp }(z)=\Delta\tanh(z/\xi_{\Delta })$ and $\Delta_{\parallel }(z)=\Delta$, where $\Delta$ is the superfluid gap in the bulk $^3$He-B.
Under the spatial dependent order parameter with $\xi_{\Delta }=\hbar v_{\rm F}/\Delta$, low energy quasiparticles with excitation energy $E_{\bm k}=\pm\Delta\sin\theta$ emerge as well as the surface bound state assuming the uniform order parameter~\cite{schung:2009,snagato:2009,swu:2013}.
The eigenfunction $\Psi_{\bm k}^{+(-)}({\bm r})$ for the positive (negative) eigenvalue is
\begin{align}
\Psi_{\bm k}^{\pm }({\bm r})\propto{\rm sech}\left(\frac{z}{2\xi }\right)e^{i{\bm k}\cdot{\bm r}}(e^{-i\phi/2}\Phi_{\uparrow }\mp e^{i\phi/2}\Phi_{\downarrow }),
\label{eq:Majorana}
\end{align}
where ${\bm k}=k_{\rm F}(\cos\phi\sin\theta,\sin\phi\sin\theta,\cos\theta)$, the coherence length is defined by $\xi\equiv\hbar v_{\rm F}/2\Delta$, and
\begin{align}
\Phi_{\uparrow }\equiv\frac{1}{\sqrt{2}}
\begin{pmatrix}
1 \\ 0 \\ 0 \\ i
\end{pmatrix},\
\Phi_{\downarrow }\equiv\frac{1}{\sqrt{2}}
\begin{pmatrix}
0 \\ i \\ -1 \\ 0
\end{pmatrix}.
\end{align}
The eigenfunction for $k_z>0$ implies the wave function of the quasiparticles reflected from the surface and that for $k_z<0$ implies the wave function of the quasiparticles injected into the surface.

The wave function of the surface bound state is produced by the wave functions of the reflected and injected quasiparticles in order to satisfy the boundary condition $\Psi=0$ at $z=0$ as
\begin{multline}
\Psi_{{\bm k}_{\parallel }}^{\pm }({\bm r})\propto{\rm sech}\left(\frac{z}{2\xi }\right)\sin(k_{\perp }z)e^{i{\bm k}_{\parallel }\cdot{\bm r}}\\
\times(e^{-i\phi/2}\Phi_{\uparrow }\mp e^{i\phi/2}\Phi_{\downarrow }),
\end{multline}
where ${\bm k}_{\parallel }$ is the parallel component to the surface and $k_{\perp }\equiv|k_z|=\sqrt{k_{\rm F}^2-k_{\parallel }^2}$.
Ignoring the gapped modes, the fermion field operator $\hat{\Psi }({\bm r})=(\hat{u}_{\uparrow }({\bm r}),\hat{u}_{\downarrow }({\bm r}),\hat{v}_{\downarrow }({\bm r}),\hat{v}_{\uparrow }({\bm r}))^{\rm T}$ is described by using the wave function as~\cite{schung:2009,snagato:2009}
\begin{align}
\hat{\Psi }({\bm r})=\sum_{{\bm k}_{\parallel }}[\hat{\gamma }_{{\bm k}_{\parallel }}\Psi_{{\bm k}_{\parallel }}^+({\bm r})+\hat{\gamma }_{-{\bm k}_{\parallel }}^{\dagger }\Psi_{{\bm k}_{\parallel }}^-({\bm r})],
\end{align}
which gives the Majorana relation $\hat{u}_{\sigma }({\bm r})=-i\hat{v}_{\sigma }({\bm r})$.
Therefore, we can regard the quasiparticles in the surface bound state as the Majorana fermions.

When we consider the scattering process on the Majorana fermions by an impurity with a small potential radius $R\ll\xi$, such as the electron bubble~\cite{sshevtsov:2016,sshevtsov:arXiv}, at $z_0$, the spatial modulation with the scale $\xi$ of the wave function in Eq.~\eqref{eq:Majorana} can be neglected as
\begin{align}
\Psi_{\bm k}^{\pm }({\bm r})\propto{\rm sech}\left(\frac{z_0}{2\xi }\right)e^{i{\bm k}\cdot{\bm r}}(e^{-i\phi/2}\Phi_{\uparrow }\mp e^{i\phi/2}\Phi_{\downarrow }).
\end{align}
Therefore, the normalized spinor for the Majorana fermions in each spin state are described by
\begin{align}
|\Psi_{{\bm k},\uparrow }^{\pm }\rangle=e^{-i\phi/2}\Phi_{\uparrow }|{\bm k}\rangle,\
|\Psi_{{\bm k},\downarrow }^{\pm }\rangle=\mp e^{i\phi/2}\Phi_{\downarrow }|{\bm k}\rangle,
\end{align}
where $\langle{\bm r}|{\bm k}\rangle=e^{i{\bm k}\cdot{\bm r}}$ indicates the plane wave.

\begin{widetext}
\section{Quasiclassical Green's function for surface bound state}

The quasiclassical Green's function in the Matsubara representation for the surface state of $^3$He-B with $\Delta_{\perp }(z)=\Delta\tanh(z/\xi_{\Delta })$ and $\Delta_{\parallel }(z)=\Delta$ is obtained in Ref.~\onlinecite{stsutsumi:2012c}.
By analytic continuation of the Matsubara Green's function, the retarded Green's function for the surface bound state $|E|<\Delta$ is derived as
\begin{align}
g_{\rm S}(\hat{\bm k},E,z)=-i\pi
\begin{pmatrix}
g_0(\theta,E,z) & -g_{\parallel }(\theta,E,z)e^{-i\phi } & -f_{\parallel }(\theta,E,z)e^{-i\phi } & f_z(\theta,E,z) \\
g_{\parallel }(\theta,E,z)e^{i\phi } & g_0(\theta,E,z) & f_z(\theta,E,z) & f_{\parallel }(\theta,E,z)e^{i\phi } \\
-f_{\parallel }(\theta,E,z)e^{i\phi } & \underline{f}_z(\theta,E,z) & g_0(\theta,E,z) & g_{\parallel }(\theta,E,z)e^{i\phi } \\
\underline{f}_z(\theta,E,z) & f_{\parallel }(\theta,E,z)e^{-i\phi } & -g_{\parallel }(\theta,E,z)e^{-i\phi } & g_0(\theta,E,z)
\end{pmatrix},
\end{align}
where $\hat{\bm k}=(\cos\phi\sin\theta,\sin\phi\sin\theta,\cos\theta)$.
Each component is given by
\begin{equation}
\begin{split}
&g_0(\theta,E,z)=-i\frac{E}{\sqrt{\Delta^2-E^2}}\left[1-\frac{1}{2}P(\theta,E)\ {\rm sech}^2\left(\frac{z}{2\xi }\right)\right],\\
&g_{\parallel }(\theta,E,z)=\frac{1}{2}\frac{\Delta\sin\theta }{\sqrt{\Delta^2-E^2}}P(\theta,E)\ {\rm sech}^2\left(\frac{z}{2\xi }\right),\\
&f_{\parallel }(\theta,E,z)=i\frac{\Delta\sin\theta }{\sqrt{\Delta^2-E^2}}\left[1-\frac{1}{2}P(\theta,E)\ {\rm sech}^2\left(\frac{z}{2\xi }\right)\right],\\
&f_z(\theta,E,z)=i\frac{\Delta\cos\theta }{\sqrt{\Delta^2-E^2}}\tanh\left(\frac{z}{2\xi }\right)+\frac{1}{2}\frac{E}{\sqrt{\Delta^2-E^2}}P(\theta,E)\ {\rm sech}^2\left(\frac{z}{2\xi }\right),\\
&\underline{f}_z(\theta,E,z)=i\frac{\Delta\cos\theta }{\sqrt{\Delta^2-E^2}}\tanh\left(\frac{z}{2\xi }\right)-\frac{1}{2}\frac{E}{\sqrt{\Delta^2-E^2}}P(\theta,E)\ {\rm sech}^2\left(\frac{z}{2\xi }\right),
\end{split}
\end{equation}
where $P(\theta,E)=\Delta^2\cos^2\theta/[(\epsilon^{\rm R})^2-\Delta^2\sin^2\theta]$ with $\epsilon^{\rm R}\equiv E+i0^+$.
\end{widetext}


\begin{thebibliography}{41}%
\makeatletter
\providecommand \@ifxundefined [1]{%
 \@ifx{#1\undefined}
}%
\providecommand \@ifnum [1]{%
 \ifnum #1\expandafter \@firstoftwo
 \else \expandafter \@secondoftwo
 \fi
}%
\providecommand \@ifx [1]{%
 \ifx #1\expandafter \@firstoftwo
 \else \expandafter \@secondoftwo
 \fi
}%
\providecommand \natexlab [1]{#1}%
\providecommand \enquote  [1]{``#1''}%
\providecommand \bibnamefont  [1]{#1}%
\providecommand \bibfnamefont [1]{#1}%
\providecommand \citenamefont [1]{#1}%
\providecommand \href@noop [0]{\@secondoftwo}%
\providecommand \href [0]{\begingroup \@sanitize@url \@href}%
\providecommand \@href[1]{\@@startlink{#1}\@@href}%
\providecommand \@@href[1]{\endgroup#1\@@endlink}%
\providecommand \@sanitize@url [0]{\catcode `\\12\catcode `\$12\catcode
  `\&12\catcode `\#12\catcode `\^12\catcode `\_12\catcode `\%12\relax}%
\providecommand \@@startlink[1]{}%
\providecommand \@@endlink[0]{}%
\providecommand \url  [0]{\begingroup\@sanitize@url \@url }%
\providecommand \@url [1]{\endgroup\@href {#1}{\urlprefix }}%
\providecommand \urlprefix  [0]{URL }%
\providecommand \Eprint [0]{\href }%
\providecommand \doibase [0]{http://dx.doi.org/}%
\providecommand \selectlanguage [0]{\@gobble}%
\providecommand \bibinfo  [0]{\@secondoftwo}%
\providecommand \bibfield  [0]{\@secondoftwo}%
\providecommand \translation [1]{[#1]}%
\providecommand \BibitemOpen [0]{}%
\providecommand \bibitemStop [0]{}%
\providecommand \bibitemNoStop [0]{.\EOS\space}%
\providecommand \EOS [0]{\spacefactor3000\relax}%
\providecommand \BibitemShut  [1]{\csname bibitem#1\endcsname}%
\let\auto@bib@innerbib\@empty
\bibitem [{\citenamefont {Hasan}\ and\ \citenamefont
  {Kane}(2010)}]{hasan:2010}%
  \BibitemOpen
  \bibfield  {author} {\bibinfo {author} {\bibfnamefont {M.~Z.}\ \bibnamefont
  {Hasan}}\ and\ \bibinfo {author} {\bibfnamefont {C.~L.}\ \bibnamefont
  {Kane}},\ }\href@noop {} {\bibfield  {journal} {\bibinfo  {journal} {Rev.
  Mod. Phys.}\ }\textbf {\bibinfo {volume} {82}},\ \bibinfo {pages} {3045}
  (\bibinfo {year} {2010})}\BibitemShut {NoStop}%
\bibitem [{\citenamefont {Qi}\ and\ \citenamefont {Zhang}(2011)}]{qi:2011}%
  \BibitemOpen
  \bibfield  {author} {\bibinfo {author} {\bibfnamefont {X.-L.}\ \bibnamefont
  {Qi}}\ and\ \bibinfo {author} {\bibfnamefont {S.-C.}\ \bibnamefont {Zhang}},\
  }\href@noop {} {\bibfield  {journal} {\bibinfo  {journal} {Rev. Mod. Phys.}\
  }\textbf {\bibinfo {volume} {83}},\ \bibinfo {pages} {1057} (\bibinfo {year}
  {2011})}\BibitemShut {NoStop}%
\bibitem [{\citenamefont {Mizushima}\ \emph {et~al.}(2015)\citenamefont
  {Mizushima}, \citenamefont {Tsutsumi}, \citenamefont {Sato},\ and\
  \citenamefont {Machida}}]{mizushima:2015}%
  \BibitemOpen
  \bibfield  {author} {\bibinfo {author} {\bibfnamefont {T.}~\bibnamefont
  {Mizushima}}, \bibinfo {author} {\bibfnamefont {Y.}~\bibnamefont {Tsutsumi}},
  \bibinfo {author} {\bibfnamefont {M.}~\bibnamefont {Sato}}, \ and\ \bibinfo
  {author} {\bibfnamefont {K.}~\bibnamefont {Machida}},\ }\href@noop {}
  {\bibfield  {journal} {\bibinfo  {journal} {J. Phys.: Condens. Matter}\
  }\textbf {\bibinfo {volume} {27}},\ \bibinfo {pages} {113203} (\bibinfo
  {year} {2015})}\BibitemShut {NoStop}%
\bibitem [{\citenamefont {Mizushima}\ \emph {et~al.}(2016)\citenamefont
  {Mizushima}, \citenamefont {Tsutsumi}, \citenamefont {Kawakami},
  \citenamefont {Sato}, \citenamefont {Ichioka},\ and\ \citenamefont
  {Machida}}]{mizushima:2016}%
  \BibitemOpen
  \bibfield  {author} {\bibinfo {author} {\bibfnamefont {T.}~\bibnamefont
  {Mizushima}}, \bibinfo {author} {\bibfnamefont {Y.}~\bibnamefont {Tsutsumi}},
  \bibinfo {author} {\bibfnamefont {T.}~\bibnamefont {Kawakami}}, \bibinfo
  {author} {\bibfnamefont {M.}~\bibnamefont {Sato}}, \bibinfo {author}
  {\bibfnamefont {M.}~\bibnamefont {Ichioka}}, \ and\ \bibinfo {author}
  {\bibfnamefont {K.}~\bibnamefont {Machida}},\ }\href@noop {} {\bibfield
  {journal} {\bibinfo  {journal} {J. Phys. Soc. Jpn.}\ }\textbf {\bibinfo
  {volume} {85}},\ \bibinfo {pages} {022001} (\bibinfo {year}
  {2016})}\BibitemShut {NoStop}%
\bibitem [{\citenamefont {Fu}\ and\ \citenamefont {Kane}(2008)}]{fu:2008}%
  \BibitemOpen
  \bibfield  {author} {\bibinfo {author} {\bibfnamefont {L.}~\bibnamefont
  {Fu}}\ and\ \bibinfo {author} {\bibfnamefont {C.~L.}\ \bibnamefont {Kane}},\
  }\href@noop {} {\bibfield  {journal} {\bibinfo  {journal} {Phys. Rev. Lett.}\
  }\textbf {\bibinfo {volume} {100}},\ \bibinfo {pages} {096407} (\bibinfo
  {year} {2008})}\BibitemShut {NoStop}%
\bibitem [{\citenamefont {Schnyder}\ \emph {et~al.}(2008)\citenamefont
  {Schnyder}, \citenamefont {Ryu}, \citenamefont {Furusaki},\ and\
  \citenamefont {Ludwig}}]{schnyder:2008}%
  \BibitemOpen
  \bibfield  {author} {\bibinfo {author} {\bibfnamefont {A.~P.}\ \bibnamefont
  {Schnyder}}, \bibinfo {author} {\bibfnamefont {S.}~\bibnamefont {Ryu}},
  \bibinfo {author} {\bibfnamefont {A.}~\bibnamefont {Furusaki}}, \ and\
  \bibinfo {author} {\bibfnamefont {A.~W.~W.}\ \bibnamefont {Ludwig}},\
  }\href@noop {} {\bibfield  {journal} {\bibinfo  {journal} {Phys. Rev. B}\
  }\textbf {\bibinfo {volume} {78}},\ \bibinfo {pages} {195125} (\bibinfo
  {year} {2008})}\BibitemShut {NoStop}%
\bibitem [{\citenamefont {Sato}\ \emph {et~al.}(2009)\citenamefont {Sato},
  \citenamefont {Takahashi},\ and\ \citenamefont {Fujimoto}}]{sato:2009}%
  \BibitemOpen
  \bibfield  {author} {\bibinfo {author} {\bibfnamefont {M.}~\bibnamefont
  {Sato}}, \bibinfo {author} {\bibfnamefont {Y.}~\bibnamefont {Takahashi}}, \
  and\ \bibinfo {author} {\bibfnamefont {S.}~\bibnamefont {Fujimoto}},\
  }\href@noop {} {\bibfield  {journal} {\bibinfo  {journal} {Phys. Rev. Lett.}\
  }\textbf {\bibinfo {volume} {103}},\ \bibinfo {pages} {020401} (\bibinfo
  {year} {2009})}\BibitemShut {NoStop}%
\bibitem [{\citenamefont {Lutchyn}\ \emph {et~al.}(2010)\citenamefont
  {Lutchyn}, \citenamefont {Sau},\ and\ \citenamefont
  {Das~Sarma}}]{lutchyn:2010}%
  \BibitemOpen
  \bibfield  {author} {\bibinfo {author} {\bibfnamefont {R.~M.}\ \bibnamefont
  {Lutchyn}}, \bibinfo {author} {\bibfnamefont {J.~D.}\ \bibnamefont {Sau}}, \
  and\ \bibinfo {author} {\bibfnamefont {S.}~\bibnamefont {Das~Sarma}},\
  }\href@noop {} {\bibfield  {journal} {\bibinfo  {journal} {Phys. Rev. Lett.}\
  }\textbf {\bibinfo {volume} {105}},\ \bibinfo {pages} {077001} (\bibinfo
  {year} {2010})}\BibitemShut {NoStop}%
\bibitem [{\citenamefont {Sasaki}\ \emph {et~al.}(2011)\citenamefont {Sasaki},
  \citenamefont {Kriener}, \citenamefont {Segawa}, \citenamefont {Yada},
  \citenamefont {Tanaka}, \citenamefont {Sato},\ and\ \citenamefont
  {Ando}}]{sasaki:2011}%
  \BibitemOpen
  \bibfield  {author} {\bibinfo {author} {\bibfnamefont {S.}~\bibnamefont
  {Sasaki}}, \bibinfo {author} {\bibfnamefont {M.}~\bibnamefont {Kriener}},
  \bibinfo {author} {\bibfnamefont {K.}~\bibnamefont {Segawa}}, \bibinfo
  {author} {\bibfnamefont {K.}~\bibnamefont {Yada}}, \bibinfo {author}
  {\bibfnamefont {Y.}~\bibnamefont {Tanaka}}, \bibinfo {author} {\bibfnamefont
  {M.}~\bibnamefont {Sato}}, \ and\ \bibinfo {author} {\bibfnamefont
  {Y.}~\bibnamefont {Ando}},\ }\href@noop {} {\bibfield  {journal} {\bibinfo
  {journal} {Phys. Rev. Lett.}\ }\textbf {\bibinfo {volume} {107}},\ \bibinfo
  {pages} {217001} (\bibinfo {year} {2011})}\BibitemShut {NoStop}%
\bibitem [{\citenamefont {Tsutsumi}\ \emph {et~al.}(2013)\citenamefont
  {Tsutsumi}, \citenamefont {Ishikawa}, \citenamefont {Kawakami}, \citenamefont
  {Mizushima}, \citenamefont {Sato}, \citenamefont {Ichioka},\ and\
  \citenamefont {Machida}}]{tsutsumi:2013}%
  \BibitemOpen
  \bibfield  {author} {\bibinfo {author} {\bibfnamefont {Y.}~\bibnamefont
  {Tsutsumi}}, \bibinfo {author} {\bibfnamefont {M.}~\bibnamefont {Ishikawa}},
  \bibinfo {author} {\bibfnamefont {T.}~\bibnamefont {Kawakami}}, \bibinfo
  {author} {\bibfnamefont {T.}~\bibnamefont {Mizushima}}, \bibinfo {author}
  {\bibfnamefont {M.}~\bibnamefont {Sato}}, \bibinfo {author} {\bibfnamefont
  {M.}~\bibnamefont {Ichioka}}, \ and\ \bibinfo {author} {\bibfnamefont
  {K.}~\bibnamefont {Machida}},\ }\href@noop {} {\bibfield  {journal} {\bibinfo
   {journal} {J. Phys. Soc. Jpn.}\ }\textbf {\bibinfo {volume} {82}},\ \bibinfo
  {pages} {113707} (\bibinfo {year} {2013})}\BibitemShut {NoStop}%
\bibitem [{\citenamefont {Tsutsumi}\ \emph {et~al.}(2015)\citenamefont
  {Tsutsumi}, \citenamefont {Kawakami}, \citenamefont {Shiozaki}, \citenamefont
  {Sato},\ and\ \citenamefont {Machida}}]{tsutsumi:2015}%
  \BibitemOpen
  \bibfield  {author} {\bibinfo {author} {\bibfnamefont {Y.}~\bibnamefont
  {Tsutsumi}}, \bibinfo {author} {\bibfnamefont {T.}~\bibnamefont {Kawakami}},
  \bibinfo {author} {\bibfnamefont {K.}~\bibnamefont {Shiozaki}}, \bibinfo
  {author} {\bibfnamefont {M.}~\bibnamefont {Sato}}, \ and\ \bibinfo {author}
  {\bibfnamefont {K.}~\bibnamefont {Machida}},\ }\href@noop {} {\bibfield
  {journal} {\bibinfo  {journal} {Phys. Rev. B}\ }\textbf {\bibinfo {volume}
  {91}},\ \bibinfo {pages} {144504} (\bibinfo {year} {2015})}\BibitemShut
  {NoStop}%
\bibitem [{\citenamefont {Xu}\ \emph {et~al.}(2015)\citenamefont {Xu},
  \citenamefont {Wang}, \citenamefont {Liu}, \citenamefont {Ge}, \citenamefont
  {Yang}, \citenamefont {Liu}, \citenamefont {Xu}, \citenamefont {Guan},
  \citenamefont {Gao}, \citenamefont {Qian}, \citenamefont {Liu}, \citenamefont
  {Wang}, \citenamefont {Zhang}, \citenamefont {Xue},\ and\ \citenamefont
  {Jia}}]{xu:2015}%
  \BibitemOpen
  \bibfield  {author} {\bibinfo {author} {\bibfnamefont {J.-P.}\ \bibnamefont
  {Xu}}, \bibinfo {author} {\bibfnamefont {M.-X.}\ \bibnamefont {Wang}},
  \bibinfo {author} {\bibfnamefont {Z.~L.}\ \bibnamefont {Liu}}, \bibinfo
  {author} {\bibfnamefont {J.-F.}\ \bibnamefont {Ge}}, \bibinfo {author}
  {\bibfnamefont {X.}~\bibnamefont {Yang}}, \bibinfo {author} {\bibfnamefont
  {C.}~\bibnamefont {Liu}}, \bibinfo {author} {\bibfnamefont {Z.~A.}\
  \bibnamefont {Xu}}, \bibinfo {author} {\bibfnamefont {D.}~\bibnamefont
  {Guan}}, \bibinfo {author} {\bibfnamefont {C.~L.}\ \bibnamefont {Gao}},
  \bibinfo {author} {\bibfnamefont {D.}~\bibnamefont {Qian}}, \bibinfo {author}
  {\bibfnamefont {Y.}~\bibnamefont {Liu}}, \bibinfo {author} {\bibfnamefont
  {Q.-H.}\ \bibnamefont {Wang}}, \bibinfo {author} {\bibfnamefont {F.-C.}\
  \bibnamefont {Zhang}}, \bibinfo {author} {\bibfnamefont {Q.-K.}\ \bibnamefont
  {Xue}}, \ and\ \bibinfo {author} {\bibfnamefont {J.-F.}\ \bibnamefont
  {Jia}},\ }\href@noop {} {\bibfield  {journal} {\bibinfo  {journal} {Phys.
  Rev. Lett.}\ }\textbf {\bibinfo {volume} {114}},\ \bibinfo {pages} {017001}
  (\bibinfo {year} {2015})}\BibitemShut {NoStop}%
\bibitem [{\citenamefont {Kawakami}\ and\ \citenamefont
  {Hu}(2015)}]{kawakami:2015}%
  \BibitemOpen
  \bibfield  {author} {\bibinfo {author} {\bibfnamefont {T.}~\bibnamefont
  {Kawakami}}\ and\ \bibinfo {author} {\bibfnamefont {X.}~\bibnamefont {Hu}},\
  }\href@noop {} {\bibfield  {journal} {\bibinfo  {journal} {Phys. Rev. Lett.}\
  }\textbf {\bibinfo {volume} {115}},\ \bibinfo {pages} {177001} (\bibinfo
  {year} {2015})}\BibitemShut {NoStop}%
\bibitem [{\citenamefont {Vollhardt}\ and\ \citenamefont
  {W{\"o}lfle}(1990)}]{vollhardt:book}%
  \BibitemOpen
  \bibfield  {author} {\bibinfo {author} {\bibfnamefont {D.}~\bibnamefont
  {Vollhardt}}\ and\ \bibinfo {author} {\bibfnamefont {P.}~\bibnamefont
  {W{\"o}lfle}},\ }\href@noop {} {\emph {\bibinfo {title} {The Superfluid Phase
  of Helium 3}}}\ (\bibinfo  {publisher} {Taylor and Francis},\ \bibinfo
  {address} {London},\ \bibinfo {year} {1990})\BibitemShut {NoStop}%
\bibitem [{\citenamefont {Volovik}(2003)}]{volovik:book}%
  \BibitemOpen
  \bibfield  {author} {\bibinfo {author} {\bibfnamefont {G.~E.}\ \bibnamefont
  {Volovik}},\ }\href@noop {} {\emph {\bibinfo {title} {The Universe in a
  Helium Droplet}}}\ (\bibinfo  {publisher} {Clarendon},\ \bibinfo {address}
  {Oxford},\ \bibinfo {year} {2003})\BibitemShut {NoStop}%
\bibitem [{\citenamefont {Chung}\ and\ \citenamefont
  {Zhang}(2009)}]{chung:2009}%
  \BibitemOpen
  \bibfield  {author} {\bibinfo {author} {\bibfnamefont {S.~B.}\ \bibnamefont
  {Chung}}\ and\ \bibinfo {author} {\bibfnamefont {S.-C.}\ \bibnamefont
  {Zhang}},\ }\href@noop {} {\bibfield  {journal} {\bibinfo  {journal} {Phys.
  Rev. Lett.}\ }\textbf {\bibinfo {volume} {103}},\ \bibinfo {pages} {235301}
  (\bibinfo {year} {2009})}\BibitemShut {NoStop}%
\bibitem [{\citenamefont {Nagato}\ \emph {et~al.}(2009)\citenamefont {Nagato},
  \citenamefont {Higashitani},\ and\ \citenamefont {Nagai}}]{nagato:2009}%
  \BibitemOpen
  \bibfield  {author} {\bibinfo {author} {\bibfnamefont {Y.}~\bibnamefont
  {Nagato}}, \bibinfo {author} {\bibfnamefont {S.}~\bibnamefont {Higashitani}},
  \ and\ \bibinfo {author} {\bibfnamefont {K.}~\bibnamefont {Nagai}},\
  }\href@noop {} {\bibfield  {journal} {\bibinfo  {journal} {J. Phys. Soc.
  Jpn.}\ }\textbf {\bibinfo {volume} {78}},\ \bibinfo {pages} {123603}
  (\bibinfo {year} {2009})}\BibitemShut {NoStop}%
\bibitem [{\citenamefont {Tsutsumi}\ \emph {et~al.}(2011)\citenamefont
  {Tsutsumi}, \citenamefont {Ichioka},\ and\ \citenamefont
  {Machida}}]{tsutsumi:2011b}%
  \BibitemOpen
  \bibfield  {author} {\bibinfo {author} {\bibfnamefont {Y.}~\bibnamefont
  {Tsutsumi}}, \bibinfo {author} {\bibfnamefont {M.}~\bibnamefont {Ichioka}}, \
  and\ \bibinfo {author} {\bibfnamefont {K.}~\bibnamefont {Machida}},\
  }\href@noop {} {\bibfield  {journal} {\bibinfo  {journal} {Phys. Rev. B}\
  }\textbf {\bibinfo {volume} {83}},\ \bibinfo {pages} {094510} (\bibinfo
  {year} {2011})}\BibitemShut {NoStop}%
\bibitem [{\citenamefont {Tsutsumi}\ and\ \citenamefont
  {Machida}(2012)}]{tsutsumi:2012c}%
  \BibitemOpen
  \bibfield  {author} {\bibinfo {author} {\bibfnamefont {Y.}~\bibnamefont
  {Tsutsumi}}\ and\ \bibinfo {author} {\bibfnamefont {K.}~\bibnamefont
  {Machida}},\ }\href@noop {} {\bibfield  {journal} {\bibinfo  {journal} {J.
  Phys. Soc. Jpn.}\ }\textbf {\bibinfo {volume} {81}},\ \bibinfo {pages}
  {074607} (\bibinfo {year} {2012})}\BibitemShut {NoStop}%
\bibitem [{\citenamefont {Wu}\ and\ \citenamefont {Sauls}(2013)}]{wu:2013}%
  \BibitemOpen
  \bibfield  {author} {\bibinfo {author} {\bibfnamefont {H.}~\bibnamefont
  {Wu}}\ and\ \bibinfo {author} {\bibfnamefont {J.~A.}\ \bibnamefont {Sauls}},\
  }\href@noop {} {\bibfield  {journal} {\bibinfo  {journal} {Phys. Rev. B}\
  }\textbf {\bibinfo {volume} {88}},\ \bibinfo {pages} {184506} (\bibinfo
  {year} {2013})}\BibitemShut {NoStop}%
\bibitem [{\citenamefont {Okuda}\ and\ \citenamefont
  {Nomura}(2012)}]{okuda:2012}%
  \BibitemOpen
  \bibfield  {author} {\bibinfo {author} {\bibfnamefont {Y.}~\bibnamefont
  {Okuda}}\ and\ \bibinfo {author} {\bibfnamefont {R.}~\bibnamefont {Nomura}},\
  }\href@noop {} {\bibfield  {journal} {\bibinfo  {journal} {J. Phys.: Condens.
  Matter}\ }\textbf {\bibinfo {volume} {24}},\ \bibinfo {pages} {343201}
  (\bibinfo {year} {2012})}\BibitemShut {NoStop}%
\bibitem [{\citenamefont {Bunkov}\ and\ \citenamefont
  {Gazizulin}()}]{bunkov:arXiv}%
  \BibitemOpen
  \bibfield  {author} {\bibinfo {author} {\bibfnamefont {Y.~M.}\ \bibnamefont
  {Bunkov}}\ and\ \bibinfo {author} {\bibfnamefont {R.~R.}\ \bibnamefont
  {Gazizulin}},\ }\href@noop {} {\bibinfo  {journal} {arXiv:1504.01711;
  arXiv:1605.05203}\ }\BibitemShut {NoStop}%
\bibitem [{\citenamefont {Murakawa}\ \emph {et~al.}(2009)\citenamefont
  {Murakawa}, \citenamefont {Tamura}, \citenamefont {Wada}, \citenamefont
  {Wasai}, \citenamefont {Saitoh}, \citenamefont {Aoki}, \citenamefont
  {Nomura}, \citenamefont {Okuda}, \citenamefont {Nagato}, \citenamefont
  {Yamamoto}, \citenamefont {Higashitani},\ and\ \citenamefont
  {Nagai}}]{murakawa:2009}%
  \BibitemOpen
\bibfield  {journal} {  }\bibfield  {author} {\bibinfo {author} {\bibfnamefont
  {S.}~\bibnamefont {Murakawa}}, \bibinfo {author} {\bibfnamefont
  {Y.}~\bibnamefont {Tamura}}, \bibinfo {author} {\bibfnamefont
  {Y.}~\bibnamefont {Wada}}, \bibinfo {author} {\bibfnamefont {M.}~\bibnamefont
  {Wasai}}, \bibinfo {author} {\bibfnamefont {M.}~\bibnamefont {Saitoh}},
  \bibinfo {author} {\bibfnamefont {Y.}~\bibnamefont {Aoki}}, \bibinfo {author}
  {\bibfnamefont {R.}~\bibnamefont {Nomura}}, \bibinfo {author} {\bibfnamefont
  {Y.}~\bibnamefont {Okuda}}, \bibinfo {author} {\bibfnamefont
  {Y.}~\bibnamefont {Nagato}}, \bibinfo {author} {\bibfnamefont
  {M.}~\bibnamefont {Yamamoto}}, \bibinfo {author} {\bibfnamefont
  {S.}~\bibnamefont {Higashitani}}, \ and\ \bibinfo {author} {\bibfnamefont
  {K.}~\bibnamefont {Nagai}},\ }\href@noop {} {\bibfield  {journal} {\bibinfo
  {journal} {Phys. Rev. Lett.}\ }\textbf {\bibinfo {volume} {103}},\ \bibinfo
  {pages} {155301} (\bibinfo {year} {2009})}\BibitemShut {NoStop}%
\bibitem [{\citenamefont {Ikegami}\ \emph
  {et~al.}(2013{\natexlab{a}})\citenamefont {Ikegami}, \citenamefont
  {Bum~Chung},\ and\ \citenamefont {Kono}}]{ikegami:2013b}%
  \BibitemOpen
  \bibfield  {author} {\bibinfo {author} {\bibfnamefont {H.}~\bibnamefont
  {Ikegami}}, \bibinfo {author} {\bibfnamefont {S.}~\bibnamefont {Bum~Chung}},
  \ and\ \bibinfo {author} {\bibfnamefont {K.}~\bibnamefont {Kono}},\
  }\href@noop {} {\bibfield  {journal} {\bibinfo  {journal} {J. Phys. Soc.
  Jpn.}\ }\textbf {\bibinfo {volume} {82}},\ \bibinfo {pages} {124607}
  (\bibinfo {year} {2013}{\natexlab{a}})}\BibitemShut {NoStop}%
\bibitem [{\citenamefont {Ahonen}\ \emph {et~al.}(1978)\citenamefont {Ahonen},
  \citenamefont {Kokko}, \citenamefont {Paalanen}, \citenamefont {Richardson},
  \citenamefont {Schoepe},\ and\ \citenamefont {Takano}}]{ahonen:1978}%
  \BibitemOpen
  \bibfield  {author} {\bibinfo {author} {\bibfnamefont {A.~I.}\ \bibnamefont
  {Ahonen}}, \bibinfo {author} {\bibfnamefont {J.}~\bibnamefont {Kokko}},
  \bibinfo {author} {\bibfnamefont {M.~A.}\ \bibnamefont {Paalanen}}, \bibinfo
  {author} {\bibfnamefont {R.~C.}\ \bibnamefont {Richardson}}, \bibinfo
  {author} {\bibfnamefont {W.}~\bibnamefont {Schoepe}}, \ and\ \bibinfo
  {author} {\bibfnamefont {Y.}~\bibnamefont {Takano}},\ }\href@noop {}
  {\bibfield  {journal} {\bibinfo  {journal} {J. Low Temp. Phys.}\ }\textbf
  {\bibinfo {volume} {30}},\ \bibinfo {pages} {205} (\bibinfo {year}
  {1978})}\BibitemShut {NoStop}%
\bibitem [{\citenamefont {Baym}\ \emph {et~al.}(1977)\citenamefont {Baym},
  \citenamefont {Pethick},\ and\ \citenamefont {Salomaa}}]{baym:1977}%
  \BibitemOpen
  \bibfield  {author} {\bibinfo {author} {\bibfnamefont {G.}~\bibnamefont
  {Baym}}, \bibinfo {author} {\bibfnamefont {C.~J.}\ \bibnamefont {Pethick}}, \
  and\ \bibinfo {author} {\bibfnamefont {M.}~\bibnamefont {Salomaa}},\
  }\href@noop {} {\bibfield  {journal} {\bibinfo  {journal} {Phys. Rev. Lett.}\
  }\textbf {\bibinfo {volume} {38}},\ \bibinfo {pages} {845} (\bibinfo {year}
  {1977})}\BibitemShut {NoStop}%
\bibitem [{\citenamefont {Baym}\ \emph {et~al.}(1979)\citenamefont {Baym},
  \citenamefont {Pethick},\ and\ \citenamefont {Salomaa}}]{baym:1979}%
  \BibitemOpen
  \bibfield  {author} {\bibinfo {author} {\bibfnamefont {G.}~\bibnamefont
  {Baym}}, \bibinfo {author} {\bibfnamefont {C.}~\bibnamefont {Pethick}}, \
  and\ \bibinfo {author} {\bibfnamefont {M.}~\bibnamefont {Salomaa}},\
  }\href@noop {} {\bibfield  {journal} {\bibinfo  {journal} {J. Low Temp.
  Phys.}\ }\textbf {\bibinfo {volume} {36}},\ \bibinfo {pages} {431} (\bibinfo
  {year} {1979})}\BibitemShut {NoStop}%
\bibitem [{\citenamefont {Salomaa}\ \emph {et~al.}(1980)\citenamefont
  {Salomaa}, \citenamefont {Pethick},\ and\ \citenamefont
  {Baym}}]{salomaa:1980}%
  \BibitemOpen
  \bibfield  {author} {\bibinfo {author} {\bibfnamefont {M.}~\bibnamefont
  {Salomaa}}, \bibinfo {author} {\bibfnamefont {C.~J.}\ \bibnamefont
  {Pethick}}, \ and\ \bibinfo {author} {\bibfnamefont {G.}~\bibnamefont
  {Baym}},\ }\href@noop {} {\bibfield  {journal} {\bibinfo  {journal} {J. Low
  Temp. Phys.}\ }\textbf {\bibinfo {volume} {40}},\ \bibinfo {pages} {297}
  (\bibinfo {year} {1980})}\BibitemShut {NoStop}%
\bibitem [{\citenamefont {Ahonen}\ \emph {et~al.}(1976)\citenamefont {Ahonen},
  \citenamefont {Kokko}, \citenamefont {Lounasmaa}, \citenamefont {Paalanen},
  \citenamefont {Richardson}, \citenamefont {Schoepe},\ and\ \citenamefont
  {Takano}}]{ahonen:1976}%
  \BibitemOpen
  \bibfield  {author} {\bibinfo {author} {\bibfnamefont {A.~I.}\ \bibnamefont
  {Ahonen}}, \bibinfo {author} {\bibfnamefont {J.}~\bibnamefont {Kokko}},
  \bibinfo {author} {\bibfnamefont {O.~V.}\ \bibnamefont {Lounasmaa}}, \bibinfo
  {author} {\bibfnamefont {M.~A.}\ \bibnamefont {Paalanen}}, \bibinfo {author}
  {\bibfnamefont {R.~C.}\ \bibnamefont {Richardson}}, \bibinfo {author}
  {\bibfnamefont {W.}~\bibnamefont {Schoepe}}, \ and\ \bibinfo {author}
  {\bibfnamefont {Y.}~\bibnamefont {Takano}},\ }\href@noop {} {\bibfield
  {journal} {\bibinfo  {journal} {Phys. Rev. Lett.}\ }\textbf {\bibinfo
  {volume} {37}},\ \bibinfo {pages} {511} (\bibinfo {year} {1976})}\BibitemShut
  {NoStop}%
\bibitem [{\citenamefont {Roach}\ \emph {et~al.}(1977)\citenamefont {Roach},
  \citenamefont {Ketterson},\ and\ \citenamefont {Roach}}]{roach:1977}%
  \BibitemOpen
  \bibfield  {author} {\bibinfo {author} {\bibfnamefont {P.~D.}\ \bibnamefont
  {Roach}}, \bibinfo {author} {\bibfnamefont {J.~B.}\ \bibnamefont
  {Ketterson}}, \ and\ \bibinfo {author} {\bibfnamefont {P.~R.}\ \bibnamefont
  {Roach}},\ }\href@noop {} {\bibfield  {journal} {\bibinfo  {journal} {Phys.
  Rev. Lett.}\ }\textbf {\bibinfo {volume} {39}},\ \bibinfo {pages} {626}
  (\bibinfo {year} {1977})}\BibitemShut {NoStop}%
\bibitem [{\citenamefont {Ikegami}\ \emph
  {et~al.}(2013{\natexlab{b}})\citenamefont {Ikegami}, \citenamefont
  {Tsutsumi},\ and\ \citenamefont {Kono}}]{ikegami:2013}%
  \BibitemOpen
  \bibfield  {author} {\bibinfo {author} {\bibfnamefont {H.}~\bibnamefont
  {Ikegami}}, \bibinfo {author} {\bibfnamefont {Y.}~\bibnamefont {Tsutsumi}}, \
  and\ \bibinfo {author} {\bibfnamefont {K.}~\bibnamefont {Kono}},\ }\href@noop
  {} {\bibfield  {journal} {\bibinfo  {journal} {Science}\ }\textbf {\bibinfo
  {volume} {341}},\ \bibinfo {pages} {59} (\bibinfo {year}
  {2013}{\natexlab{b}})}\BibitemShut {NoStop}%
\bibitem [{\citenamefont {Ikegami}\ \emph {et~al.}(2015)\citenamefont
  {Ikegami}, \citenamefont {Tsutsumi},\ and\ \citenamefont
  {Kono}}]{ikegami:2015}%
  \BibitemOpen
  \bibfield  {author} {\bibinfo {author} {\bibfnamefont {H.}~\bibnamefont
  {Ikegami}}, \bibinfo {author} {\bibfnamefont {Y.}~\bibnamefont {Tsutsumi}}, \
  and\ \bibinfo {author} {\bibfnamefont {K.}~\bibnamefont {Kono}},\ }\href@noop
  {} {\bibfield  {journal} {\bibinfo  {journal} {J. Phys. Soc. Jpn.}\ }\textbf
  {\bibinfo {volume} {84}},\ \bibinfo {pages} {044602} (\bibinfo {year}
  {2015})}\BibitemShut {NoStop}%
\bibitem [{\citenamefont {Shevtsov}\ and\ \citenamefont
  {Sauls}(2016)}]{shevtsov:2016}%
  \BibitemOpen
  \bibfield  {author} {\bibinfo {author} {\bibfnamefont {O.}~\bibnamefont
  {Shevtsov}}\ and\ \bibinfo {author} {\bibfnamefont {J.~A.}\ \bibnamefont
  {Sauls}},\ }\href@noop {} {\bibfield  {journal} {\bibinfo  {journal} {Phys.
  Rev. B}\ }\textbf {\bibinfo {volume} {94}},\ \bibinfo {pages} {064511}
  (\bibinfo {year} {2016})}\BibitemShut {NoStop}%
\bibitem [{\citenamefont {Bromley}(1981)}]{bromley:1981}%
  \BibitemOpen
\bibfield  {journal} {  }\bibfield  {author} {\bibinfo {author} {\bibfnamefont
  {D.~J.}\ \bibnamefont {Bromley}},\ }\href@noop {} {\bibfield  {journal}
  {\bibinfo  {journal} {Phys. Rev. B}\ }\textbf {\bibinfo {volume} {23}},\
  \bibinfo {pages} {4503} (\bibinfo {year} {1981})}\BibitemShut {NoStop}%
\bibitem [{sup({\natexlab{a}})}]{supplement1}%
  \BibitemOpen
  \href@noop {} {\bibfield  {journal} {\bibinfo  {journal} {See Supplemental
  Material for Majorana fermions in the surface bound state, which includes Refs.~\cite{kawakami:2011,matsumoto:2001}}}}\BibitemShut {NoStop}%
\bibitem [{\citenamefont {Kawakami}\ \emph {et~al.}(2011)\citenamefont
  {Kawakami}, \citenamefont {Mizushima},\ and\ \citenamefont
  {Machida}}]{kawakami:2011}%
  \BibitemOpen
  \bibfield  {author} {\bibinfo {author} {\bibfnamefont {T.}~\bibnamefont
  {Kawakami}}, \bibinfo {author} {\bibfnamefont {T.}~\bibnamefont {Mizushima}},
  \ and\ \bibinfo {author} {\bibfnamefont {K.}~\bibnamefont {Machida}},\ }\href@noop {}
  {\bibfield  {journal} {\bibinfo  {journal}
  {J. Phys. Soc. Jpn.}\ }\textbf {\bibinfo {volume} {80}},\ \bibinfo {pages}
  {044603} (\bibinfo {year} {2011})}\BibitemShut {NoStop}%
\bibitem [{\citenamefont {Matsumoto}\ and\ \citenamefont
  {Heeb}(2001)}]{matsumoto:2001}%
  \BibitemOpen
  \bibfield  {author} {\bibinfo {author} {\bibfnamefont {M.}~\bibnamefont
  {Matsumoto}}\ and\ \bibinfo {author} {\bibfnamefont {R.}~\bibnamefont
  {Heeb}},\ }\href@noop {} {\bibfield  {journal}
  {\bibinfo  {journal} {Phys. Rev. B}\ }\textbf {\bibinfo {volume} {65}},\
  \bibinfo {pages} {014504} (\bibinfo {year} {2001})}\BibitemShut {NoStop}%
\bibitem [{\citenamefont {Thuneberg}\ \emph {et~al.}(1981)\citenamefont
  {Thuneberg}, \citenamefont {Kurkijarvi},\ and\ \citenamefont
  {Rainer}}]{thuneberg:1981}%
  \BibitemOpen
  \bibfield  {author} {\bibinfo {author} {\bibfnamefont {E.~V.}\ \bibnamefont
  {Thuneberg}}, \bibinfo {author} {\bibfnamefont {J.}~\bibnamefont
  {Kurkijarvi}}, \ and\ \bibinfo {author} {\bibfnamefont {D.}~\bibnamefont
  {Rainer}},\ }\href@noop {} {\bibfield  {journal} {\bibinfo  {journal} {J.
  Phys. C: Solid State Phys.}\ }\textbf {\bibinfo {volume} {14}},\ \bibinfo
  {pages} {5615} (\bibinfo {year} {1981})}\BibitemShut {NoStop}%
\bibitem [{sup({\natexlab{b}})}]{supplement2}%
  \BibitemOpen
  \href@noop {} {\bibfield  {journal} {\bibinfo  {journal} {See Supplemental
  Material for quasiclassical Green's function for the surface bound
  state}}}\BibitemShut {NoStop}%
  \bibitem [{\citenamefont {Shevtsov}\ and\ \citenamefont
  {Sauls}()}]{shevtsov:arXiv}%
  \BibitemOpen
  \bibfield  {author} {\bibinfo {author} {\bibfnamefont {O.}~\bibnamefont
  {Shevtsov}}\ and\ \bibinfo {author} {\bibfnamefont {J.~A.}\ \bibnamefont
  {Sauls}},\ }\href@noop {} {\bibinfo  {journal} {arXiv:1608.01644}\
  }\BibitemShut {NoStop}%
\bibitem{ikegami:private}
H. Ikegami (private communication).
\end{thebibliography}

\begin{thebibliography}{41}%
\makeatletter
\providecommand \@ifxundefined [1]{%
 \@ifx{#1\undefined}
}%
\providecommand \@ifnum [1]{%
 \ifnum #1\expandafter \@firstoftwo
 \else \expandafter \@secondoftwo
 \fi
}%
\providecommand \@ifx [1]{%
 \ifx #1\expandafter \@firstoftwo
 \else \expandafter \@secondoftwo
 \fi
}%
\providecommand \natexlab [1]{#1}%
\providecommand \enquote  [1]{``#1''}%
\providecommand \bibnamefont  [1]{#1}%
\providecommand \bibfnamefont [1]{#1}%
\providecommand \citenamefont [1]{#1}%
\providecommand \href@noop [0]{\@secondoftwo}%
\providecommand \href [0]{\begingroup \@sanitize@url \@href}%
\providecommand \@href[1]{\@@startlink{#1}\@@href}%
\providecommand \@@href[1]{\endgroup#1\@@endlink}%
\providecommand \@sanitize@url [0]{\catcode `\\12\catcode `\$12\catcode
  `\&12\catcode `\#12\catcode `\^12\catcode `\_12\catcode `\%12\relax}%
\providecommand \@@startlink[1]{}%
\providecommand \@@endlink[0]{}%
\providecommand \url  [0]{\begingroup\@sanitize@url \@url }%
\providecommand \@url [1]{\endgroup\@href {#1}{\urlprefix }}%
\providecommand \urlprefix  [0]{URL }%
\providecommand \Eprint [0]{\href }%
\providecommand \doibase [0]{http://dx.doi.org/}%
\providecommand \selectlanguage [0]{\@gobble}%
\providecommand \bibinfo  [0]{\@secondoftwo}%
\providecommand \bibfield  [0]{\@secondoftwo}%
\providecommand \translation [1]{[#1]}%
\providecommand \BibitemOpen [0]{}%
\providecommand \bibitemStop [0]{}%
\providecommand \bibitemNoStop [0]{.\EOS\space}%
\providecommand \EOS [0]{\spacefactor3000\relax}%
\providecommand \BibitemShut  [1]{\csname bibitem#1\endcsname}%
\let\auto@bib@innerbib\@empty
\bibitem [{\citenamefont {Kawakami}\ \emph {et~al.}(2011)\citenamefont
  {Kawakami}, \citenamefont {Mizushima},\ and\ \citenamefont
  {Machida}}]{skawakami:2011}%
  \BibitemOpen
  \bibfield  {author} {\bibinfo {author} {\bibfnamefont {T.}~\bibnamefont
  {Kawakami}}, \bibinfo {author} {\bibfnamefont {T.}~\bibnamefont {Mizushima}},
  \ and\ \bibinfo {author} {\bibfnamefont {K.}~\bibnamefont {Machida}},\ }\href@noop {}
  {\bibfield  {journal} {\bibinfo  {journal}
  {J. Phys. Soc. Jpn.}\ }\textbf {\bibinfo {volume} {80}},\ \bibinfo {pages}
  {044603} (\bibinfo {year} {2011})}\BibitemShut {NoStop}%
\bibitem [{\citenamefont {Matsumoto}\ and\ \citenamefont
  {Heeb}(2001)}]{smatsumoto:2001}%
  \BibitemOpen
  \bibfield  {author} {\bibinfo {author} {\bibfnamefont {M.}~\bibnamefont
  {Matsumoto}}\ and\ \bibinfo {author} {\bibfnamefont {R.}~\bibnamefont
  {Heeb}},\ }\href@noop {} {\bibfield  {journal}
  {\bibinfo  {journal} {Phys. Rev. B}\ }\textbf {\bibinfo {volume} {65}},\
  \bibinfo {pages} {014504} (\bibinfo {year} {2001})}\BibitemShut {NoStop}%
\bibitem [{\citenamefont {Chung}\ and\ \citenamefont
  {Zhang}(2009)}]{schung:2009}%
  \BibitemOpen
  \bibfield  {author} {\bibinfo {author} {\bibfnamefont {S.~B.}\ \bibnamefont
  {Chung}}\ and\ \bibinfo {author} {\bibfnamefont {S.-C.}\ \bibnamefont
  {Zhang}},\ }\href@noop {} {\bibfield  {journal} {\bibinfo  {journal} {Phys.
  Rev. Lett.}\ }\textbf {\bibinfo {volume} {103}},\ \bibinfo {pages} {235301}
  (\bibinfo {year} {2009})}\BibitemShut {NoStop}%
\bibitem [{\citenamefont {Nagato}\ \emph {et~al.}(2009)\citenamefont {Nagato},
  \citenamefont {Higashitani},\ and\ \citenamefont {Nagai}}]{snagato:2009}%
  \BibitemOpen
  \bibfield  {author} {\bibinfo {author} {\bibfnamefont {Y.}~\bibnamefont
  {Nagato}}, \bibinfo {author} {\bibfnamefont {S.}~\bibnamefont {Higashitani}},
  \ and\ \bibinfo {author} {\bibfnamefont {K.}~\bibnamefont {Nagai}},\
  }\href@noop {} {\bibfield  {journal} {\bibinfo  {journal} {J. Phys. Soc.
  Jpn.}\ }\textbf {\bibinfo {volume} {78}},\ \bibinfo {pages} {123603}
  (\bibinfo {year} {2009})}\BibitemShut {NoStop}%
\bibitem [{\citenamefont {Wu}\ and\ \citenamefont {Sauls}(2013)}]{swu:2013}%
  \BibitemOpen
  \bibfield  {author} {\bibinfo {author} {\bibfnamefont {H.}~\bibnamefont
  {Wu}}\ and\ \bibinfo {author} {\bibfnamefont {J.~A.}\ \bibnamefont {Sauls}},\
  }\href@noop {} {\bibfield  {journal} {\bibinfo  {journal} {Phys. Rev. B}\
  }\textbf {\bibinfo {volume} {88}},\ \bibinfo {pages} {184506} (\bibinfo
  {year} {2013})}\BibitemShut {NoStop}%
\bibitem [{\citenamefont {Shevtsov}\ and\ \citenamefont
  {Sauls}(2016)}]{sshevtsov:2016}%
  \BibitemOpen
  \bibfield  {author} {\bibinfo {author} {\bibfnamefont {O.}~\bibnamefont
  {Shevtsov}}\ and\ \bibinfo {author} {\bibfnamefont {J.~A.}\ \bibnamefont
  {Sauls}},\ }\href@noop {} {\bibfield  {journal} {\bibinfo  {journal} {Phys.
  Rev. B}\ }\textbf {\bibinfo {volume} {94}},\ \bibinfo {pages} {064511}
  (\bibinfo {year} {2016})}\BibitemShut {NoStop}%
\bibitem [{\citenamefont {Shevtsov}\ and\ \citenamefont
  {Sauls}()}]{sshevtsov:arXiv}%
  \BibitemOpen
  \bibfield  {author} {\bibinfo {author} {\bibfnamefont {O.}~\bibnamefont
  {Shevtsov}}\ and\ \bibinfo {author} {\bibfnamefont {J.~A.}\ \bibnamefont
  {Sauls}},\ }\href@noop {} {\bibinfo  {journal} {arXiv:1608.01644}\
  }\BibitemShut {NoStop}%
\bibitem [{\citenamefont {Tsutsumi}\ and\ \citenamefont
  {Machida}(2012)}]{stsutsumi:2012c}%
  \BibitemOpen
  \bibfield  {author} {\bibinfo {author} {\bibfnamefont {Y.}~\bibnamefont
  {Tsutsumi}}\ and\ \bibinfo {author} {\bibfnamefont {K.}~\bibnamefont
  {Machida}},\ }\href@noop {} {\bibfield  {journal} {\bibinfo  {journal} {J.
  Phys. Soc. Jpn.}\ }\textbf {\bibinfo {volume} {81}},\ \bibinfo {pages}
  {074607} (\bibinfo {year} {2012})}\BibitemShut {NoStop}%
\end{thebibliography}

%

\end{document}